\documentclass[12pt]{article}
\usepackage{amsmath,amssymb,latexsym,graphicx,citesort}
\makeindex

\newtheorem{theorem}{Theorem}[section]
\newtheorem{lemma}[theorem]{Lemma}
\newtheorem{corollary}[theorem]{Corollary}

\newtheorem{proposition}[theorem]{Proposition}
\newtheorem{definition}[theorem]{Definition}

\newtheorem{remark}[theorem]{Remark}

\newcommand{\RR}{\mathbb{R}}

\newcommand{\ZZ}{\mathbb{Z}}
\newcommand{\CC}{\mathbb{C}}
\newcommand{\TT}{\mathbb{T}}
\newcommand{\T}{\mathbb{T}}

\newcommand{\B}{\mathbf{B}}
\newcommand{\W}{\mathbf{W}}
\newcommand{\G}{\Gamma}
\newcommand{\g}{\gamma}

\newcommand{\fh}{\hat{f}}

\newcommand{\D}{\mathcal{D}}

\newcommand{\dn}{d\,'}

\newcommand{\WS}{Wigner-Seitz }
\newcommand{\bd}{\begin{definition}}
\newcommand{\ed}{\end{definition}}
\newcommand{\bt}{\begin{theorem}}
\newcommand{\et}{\end{theorem}}
\newcommand{\bl}{\begin{lemma}}
\newcommand{\el}{\end{lemma}}
\newcommand{\bc}{\begin{corollary}}
\newcommand{\ec}{\end{corollary}}
\newcommand{\br}{\begin{remark}}
\newcommand{\er}{\end{remark}}
\newcommand{\bp}{\begin{proposition}}
\newcommand{\ep}{\end{proposition}}
\newcommand{\proof}{{\bf Proof. }}
\newcommand{\eproof}{$\Box$}
\begin{document}
\title{Tight frames of exponentially decaying Wannier functions}
\author{Peter Kuchment\\
Mathematics Department\\
Texas A\&M University\\
College Station, TX 77843-3368, USA\\
kuchment@math.tamu.edu}
\date{}
\maketitle

\begin{abstract}
Let $L$ be a Schr\"{o}dinger operator $(\frac{1}{i}\nabla-A(x))^2+V(x)$ with periodic magnetic and electric potentials $A,V$, a Maxwell operator $\nabla\times\frac{1}{\varepsilon (x)}\nabla\times$ in a periodic medium, or an arbitrary self-adjoint elliptic linear partial differential operator in $\RR^n$ with coefficients periodic with respect to a lattice $\G$. Let also $S$ be a finite part of its spectrum separated by gaps from the rest of the spectrum. We consider the old question of existence of a finite set of exponentially decaying Wannier functions $w_j(x)$ such that their $\G$-shifts $w_{j,\g}(x)=w_j(x-\g)$ for $\g\in\G$ span the whole spectral subspace corresponding to $S$ in some ``nice'' manner. It is known that a topological obstruction might exist to finding exponentially decaying $w_{j,\g}$ that form an orthonormal basis of the spectral subspace. This obstruction has the form of non-triviality of certain finite dimensional (with the dimension of the fiber equal to the number $m$ of spectral bands in $S$) analytic vector bundle $\Lambda_S$ over the $n$-dimensional torus. It was shown by G.~Nenciu in 1983 that in the presence of time reversal symmetry (which implies absence of magnetic fields), and if $S$ is a single band, the bundle is trivial and thus the desired Wannier functions do exist. In 2007, G.~Panati proved that in dimensions $n\leq 3$, even if $S$ consists of several spectral bands, the time reversal symmetry removes the obstruction as well, if one uses the so called composite Wannier functions. It has not been known what could be achieved when the bundle is non-trivial (which can occur for instance in presence of magnetic fields or for Chern insulators). Let $\tau$ be the type of the bundle $\Lambda_S$, i.e. the number of open sub-domains over which it is trivial (for the trivial bundle $\tau=1$, and $\tau$ never exceeds $2^n$, where $n$ is the dimension of the coordinate space). We show that it is always possible to find a finite number $l\leq \tau m$ (and thus $m\leq l \leq 2^n m$) of exponentially decaying composite Wannier functions $w_j$ such that their $\G$-shifts form a $1$-tight frame in the spectral subspace. Here $1$-tight frame is a redundant analog of an orthogonal basis, which appears in many applications, e.g. to signal processing and communication theory. This appears to be the best one can do when the topological obstruction is present. The number $l$ is the smallest dimension of a trivial bundle containing an equivalent copy of $\Lambda_S$. In particular, $l=m$ if and only if $\Lambda_S$ is trivial, in which case an orthonormal basis of exponentially decaying composite Wannier functions is known to exist.
\end{abstract}

\section{Introduction}
Wannier functions, along with Bloch waves, play an important role in solid state physics \cite{AM}, photonic crystal theory (see, e.g., \cite{JJWM,Kuch_photchapter} for general discussion of photonic crystals and \cite{Bush_wannier,Busch_wannier2,Busch_wannier3} for Wannier function applications), and in fact in any study that involves periodic linear elliptic equations. They are heavily involved in description of electronic properties of solids, theory of polarization, photonic crystals, numerical analysis using tight-binding approximation, etc. (see, e.g. \cite{Bush_wannier,Busch_wannier2,Busch_wannier3,AM,Nenciu,Brouder,Panati,Marzari,Kohn,KohnLutt,MarzSou,Wann,Wann_web} and references therein). As it is said in \cite{MarzSou}, strongly localized Wannier functions ``are the solid-state equivalent of ``localized molecular orbitals''..., and thus provide an insightful picture of the nature of chemical bonding.'' It is in particular important to have the Wannier functions decaying as fast as possible. That is why the question of possibility of choosing a finite number of families of exponentially decaying Wannier functions that form an orthonormal basis in the spectral subspace corresponding to an isolated part $S$ of the spectrum has been intensively considered in physics literature, at least since the paper by W.~Kohn \cite{Kohn}, who showed that this is possible in $1D$. However, the problem becomes non-trivial in $2D$ and higher dimensions. And indeed a topological obstruction arises there which is not present in $1D$. Namely, existence of such a family is known \cite{Thouless} to be equivalent to triviality of certain analytic vector fiber bundle that we will call $\Lambda_S$ (see (\ref{E:bundle})). It was shown by G.~Nenciu \cite{Nenciu} in 1983 (see also \cite{Helffer}) that in the presence of time reversal symmetry, and if $S$ is a single spectral band, the bundle is trivial and thus the desired family of Wannier functions does exist. In 2007, G.~Panati \cite{Panati} proved that in dimensions $n\leq 3$, even if $S$ consists of several spectral bands, the time reversal symmetry removes the obstruction as well. In this case, though, one should naturally resort to the so called composite, or generalized Wannier functions that correspond to a finite family of bands, rather than to a single band. The activity in this direction seems to be high and still increasing \cite{MarzSou,Wann_web}.

It has not been known what could be achieved when the bundle is non-trivial (which can happen in presence of magnetic fields or for Chern insulators). In this case, a family of Wannier functions with the described properties cannot exist. One can avoid the topological obstruction relaxing the exponential decay to the square integrability of the Wannier functions, but this amounts to a decay that is considered to be way too slow. There is not much one can improve on this decay without striking the topological obstruction (e.g., a slow decay such as summability of $L^2$ norms over the shifted copies of the Wigner-Seitz cell is already impossible, see Theorem \ref{T:slow}).

In this text, we show existence of what appears to be the best exponentially decaying Wannier family one can expect in the case of a non-trivial bundle $\Lambda_S$. In order to formulate the main result, we need to introduce some notions.

\begin{definition}\label{D:frame}
\indent
\begin{itemize}
  \item A {\bf frame} in the Hilbert space $H$ is a set of vectors $\psi_j$ such that for any vector $f\in H$ the following inequality holds:
      $$
      A\|f\|^2 \leq \sum |(f,\psi_j)|^2 \leq B\|f\|^2.
      $$
  \item The frame is said to be {\bf $A$-tight}, if $B=A$ in the inequality above. In particular, for a {\bf $1$-tight} frame one has
      $$
      \|f\|^2 = \sum |(f,\psi_j)|^2.
      $$
\end{itemize}
\end{definition}

The reader should notice that there is no condition of linear independence on the vectors forming a frame, so frames are usually redundant (and thus are not bases). 

It is known (see, for instance \cite{Larson_frame}) that $1$-tight frames are exactly orthogonal projections of orthonormal bases from a larger Hilbert space and play a substitute of such bases for redundant systems of vectors. If a $1$-tight frame is linearly independent, it in fact forms an orthogonal basis.

Frames in general, and tight frames in particular, have become recently a standard tool in signal processing and communication theory, in particular due to their numerical stability and robustness with respect to noise (e.g., \cite{Daubechies,Jorg,Casazza,Ben,Ben2,Cvet,Goyal} and references therein).

We also need one more notion:
 \begin{definition}\label{D:type} (e.g., \cite{Husemoller}) \indent
 Let $\Lambda$ be a (locally trivial) vector bundle over a connected base $T$.
 \begin{itemize}
 \item The bundle is said to be of a {\bf finite type}, if there is a finite open covering $\{U_j\}_{1\leq \j \leq \tau}$ of the base $T$ such that the bundle $\Lambda$ is trivial over each of $U_j$
 \item The \textbf{type} $\tau$ of the bundle is the minimal number of elements of such a covering.
 \item We will call the {\bf dimension} (also called the {\bf rank}) of the vector bundle $\Lambda$ the dimension of its fibers.
 \end{itemize}
 \end{definition}
 \begin{remark}\label{R:type}
 \indent
 \begin{itemize}
 \item Vector bundles over a compact base (e.g., over a torus) always have a finite type.
 \item The type is equal to $1$ if and only if the bundle is trivial.
 \item The type of any vector bundle over an $n$-dimensional torus does not exceed $2^n$.
 \end{itemize}
\end{remark}

We can now formulate the following main result of the paper:

{\bf Theorem \ref{T:wan_kuch_frame}} {\em Let $L$ be a self-adjoint elliptic $\G$-periodic operator in $\RR^n, n\geq 1$ and $S\subset\RR$ be the union of $m$ spectral bands of $L$. We assume that $S$ is separated from the rest of the spectrum. Let also $\tau$ be the type of the obstacle bundle $\Lambda_S$ over the $n$-dimensional torus. Then there exists a finite number $l\leq\tau m$ (and thus $m\leq l \leq 2^n m$) of exponentially decaying composite Wannier functions
$w_j(x)$ such that their shifts $w_{j,\g}:=w_j(x-\g),\g\in\G$ form a $1$-tight frame in the spectral subspace $H_S$ of the operator $L$. This means that for any $f(x)\in H_S$, the equality holds
\begin{equation}
    \int\limits_{\RR^n}|f(x)|^2dx=\sum\limits_{j,\g}|\int\limits_{\RR^n}f(x)\overline{w_{j,\g}(x)}dx|^2.
\end{equation}
The number $l\in [m,2^n m]$ is the smallest dimension of a trivial bundle containing an equivalent copy of $\Lambda_S$. In particular, $l=m$ if and only if $\Lambda_S$ is trivial, in which case an orthonormal basis of exponentially decaying composite Wannier functions exists.}

The number $l$ of the Wannier functions $w_j$ has to exceed the number of bands, unless there is no topological obstacle in the form of non-triviality of $\Lambda_S$. This creates a redundant system of functions, and thus the orthonormal property is not achievable. The $1$-tight frame property is the best analog of orthonormality one can get in this case. For instance, it allows the control of the $L^2$ norms in terms of the projections onto the Wannier system.

The structure of the paper is as follows: Section \ref{S:notions} surveys the main notions of the Floquet-Bloch theory that will be needed for the rest of the text (they can be found in most solid state texts or in \cite{Kuch_UMN,ReedSimon,Kuch_book,Kuch_photchapter}); in Section \ref{S:wannier} we introduce the Wannier functions and discuss their properties of interest; Section \ref{S:singleband} contains an overview of the known results concerning the single band case; Section \ref{S:multiband} moves to the composite bands and composite Wannier functions. It also contains the proof of the main result of the paper. The first half of the paper provides a self-contained survey
of known results, which is done for the benefit of the reader. This part also sets up all preliminary results and notions that are needed for the proof of the main result of the text.

\section{Main notions and preliminary results}\label{S:notions}

Let $L(x,D)$ be a bounded from below\footnote{This restriction is not essential and can be removed.} self-adjoint elliptic operator in $\RR^n$. The specific nature of the operator will be irrelevant (e.g., matrix operators, such as Dirac or Maxwell can be allowed). One can thus consider, without loss of generality, one's favorite periodic operator, e.g. the Schr\"{o}dinger operator $(\frac{1}{i}\nabla-A(x))^2+V(x)$ with real periodic magnetic and electric potentials $A,V$. Some conditions need to be imposed on the potentials to define a self-adjoint operator $L$ in $L^2(\RR^n)$ (e.g., \cite{Cycon,ReedSimon}). However, for what follows these details are insignificant, and thus one can safely assume ``sufficiently nice'' coefficients of $L$.

Let $\G$ be a lattice in $\RR^n$, i.e. the set of integer linear combinations of vectors of a basis $a_1,\dots,a_n$ in $\RR^n$. (Due to the general form of $L$ that we allow, no generality will be lost if the reader assumes that $\G$ is just the integer lattice $\ZZ^n$.) The coefficients of the operator $L$ are assumed to be periodic with respect to the shifts by vectors $\g$ of $\G$. We will be using a fundamental domain $\W$ (e.g., the \WS cell in physics literature) of the lattice $\G$, i.e. such a domain that its $\G$-shifts cover the whole space with only boundary overlap.

We denote by $\G^*$ the reciprocal (or dual) lattice to $\G$. It lives in the dual space to $\RR^n$, but if an inner product $(\cdot,\cdot)$ in $\RR^n$ is fixed, then $\G^*$ can be realized in the same space $\RR^n$ as consisting of all vectors $\kappa$ such that $(\kappa,\g)\in 2\pi\ZZ$ for all $\g\in\G$ (e.g., if $\Gamma=\ZZ^n$, then $\Gamma^*=2\pi\ZZ^n$). A fundamental domain $\B$ for $\G$ will be fixed (e.g., the first Brillouin zone in physics). The quotient $\RR^n/\G^*$ forms a torus that we will denote $\T^*$. It can also be realized as $\B$ with the boundary points identified according to the action of $\G^*$, and thus folded into a torus. In particular, $\G^*$-periodic functions on $\RR^n$ are more naturally identified with functions on $\T^*$, rather than with ones on $\B$.

The periodicity of the spectral problem $Lu=\lambda u$ with respect to the lattice group $\G$ suggests to use the group Fourier transform. This immediately leads to the well known under various names both in physics and mathematics \cite{Kuch_book,Kuch_photchapter,ReedSimon,Kuch_UMN,AM} transform, which we will call here the \textbf{Bloch-Floquet transform}:
\begin{equation}\label{E:Floquet_tr}
  f(x)\mapsto\hat{f}(k,x):=
  \sum\limits_{\g \in \G} f(x+\g) e^{-ik \cdot \g}.
\end{equation}
Here $k$ is a real (we will also need complex values of $k$) $n$-dimensional vector, which we will call \textbf{quasi-momentum} (its other common names are crystal momentum and Bloch momentum). Assuming that $f$ decays sufficiently fast (e.g., if $f$ has bounded support, or is in $L^2(\RR^n$)), there is no convergence problem\footnote{In fact, like the Fourier transform, the Bloch-Fourier transform can be extended in a distributional sense to a much wider class of functions, e.g. \cite{Kuch_book}.}. It is straightforward to check that for any (even complex) quasi-momentum $k$, the function $\hat{f}(k,x)$ is $\G^*$-periodic with respect to $k$ and of the \textbf{Bloch} (also called {\bf Floquet}) \textbf{form} with respect to $x$, which means that
\begin{equation}\label{E:Bloch}
\hat{f}(k,x)=e^{ik\cdot x}v_k(x),
\end{equation}
where $v_k(x)$ is $\G$-periodic. Thus, the values $x\in\W$ and $k\in \B$ are sufficient for determining the whole function $\hat{f}(k,x)$. In particular, one can consider $\hat{f}(k,x)$ as a function $\hat{f}(k,\cdot)$ on $\B$ with values in a space of functions on $\W$. Due to $\G^*$-periodicity, it is more natural (and important for us for what follows) to consider $\hat{f}(k,\cdot)$ as a function on the torus $\T^*$ rather than on $\B$. Formally, this means the change of variables from the quasi-momenta $k=(k_1,\dots,k_n)$ to {\bf Floquet multipliers} $z=e^{ik}:=(e^{ik\cdot a_1},\dots,e^{ik\cdot a_n})$. Then the torus $\T^*$, which is the image of the space of all real quasi-momenta, gets imbedded into the complex space $\CC^n$ as the unit torus $\{z| |z_j|=1,j=1,\dots,n\}$. The space of all complex quasi-momenta becomes the set $(\CC\setminus 0)^n\subset \CC^n$ that contains all complex vectors with non-zero components. We will also need some complex neighborhoods of the space of real quasi-momenta and of the torus $\TT^*$, defined for a given $\alpha >0$:
\begin{equation}\label{E:domain_D}
\D_\alpha = \{ k \in \CC^n|\, |\mbox{Im } k \cdot a_j| < \alpha, j=1, ...,
n\},
\end{equation}
and its image under the transform $k\mapsto z$
\begin{equation}\label{E:omega-a}
\Omega_\alpha=\{z=\left( z_1,...,z_n\right) \in
    \Omega \, | \, e^{-\alpha}<|z_j|<e^\alpha,\, j=1,...,n\}.
\end{equation}
Here $\{a_j\}$ is the basis of $\G$ mentioned before.

If $f\in L^2(\RR^n)$, the series (\ref{E:Floquet_tr}) converges in the space $L^2(\B,L^2(\W))$ of square integrable functions on $\B$ with values in $L^2(\W)$.

Another simple remark which we will need is the following: $\G$-shifts of $f$ result in multiplication by exponents in the Bloch-Floquet images. Namely, if $\omega\in\G$ and $f_\omega(x)=f(x-\omega)$, then
\begin{equation}\label{E:shift}
    \widehat{f_\omega}(k,x)=e^{-ik\cdot\omega}\widehat{f}(k,x).
\end{equation}

In the case of constant coefficient linear differential operators, plane waves are generalized eigenfunctions, and thus the Fourier transform diagonalizes the operators. In the periodic case, Bloch functions (\ref{E:Bloch}) and Bloch-Floquet transform (\ref{E:Floquet_tr}) play the roles of plane waves and Fourier transform correspondingly.

In order to formulate simple but crucial for us analogs for Bloch-Floquet transform of the standard Plancherel, inversion, and Paley-Wiener theorems for the Fourier transform, we need to introduce some notions first:
\begin{definition}\label{D:exp_space}
The space $L^2_a(\RR^n)$ consists of all functions $f \in
L^2_{loc}(\RR^n)$ such that for any $0<b<a$ the following
expression is finite:
\begin{equation}\label{E:L2_a}
  \psi_b(f):=\mathop{sup}\limits_{\g\in\G}\, \|f \|_{L^2(\mathbf{W}+\g)}e^{b|\g|}<\infty.
\end{equation}
This space is equipped with the natural topology defined by
the semi-norms $\psi_b$.
\end{definition}
So, this is the space of functions which are locally squarely integrable, and whose local $L^2$ norm decays exponentially when the domain is shifted to infinity. When we will be discussing the exponential decay of a Wannier function $w$, this will always mean exponential decay of $\|w \|_{L^2(\mathbf{W}+\g)}$ with respect to $\g$. \footnote{If the coefficients of the operator $L$ are good enough, one might improve the norm in which the decay is observed. For instance, if $L$ maps functions that locally belong to the Sobolev space $H^{n/2+\epsilon}$ (with some $\epsilon >0$) into functions that are locally in $H^{n/2-r+\epsilon}$, where $r$ is the order of the operator (usually $r=2$ in all physics applications), one can get pointwise decay.}

We will also denote by $A(\Omega, H)$, where $\Omega$ is an $n$-dimensional complex domain and $H$ is a Hilbert space, the space of all $H$-valued analytic functions on $\Omega$, equipped with the topology of uniform convergence on compacta.

We are now able to formulate the promised basic properties of the Bloch-Floquet transform (e.g., \cite{Kuch_book}):
\begin{theorem}\label{T:Planch}\indent
\begin{enumerate}
\item If $f \in L^2(\RR^n)$ and $K \subset \RR^n$ is a compact, then the
series (\ref{E:Floquet_tr}) converges in the space
$L^2(\TT^*,L^2(K))$. Moreover, the following equality (Plancherel
theorem) holds:
\begin{equation}\label{E:Planch}
  \|f\|_{_{L^2(\RR^n)}}^2=\int\limits_{\B}
  \|\fh (k,\cdot)\|_{_{L^2(\W)}}^2\dn k=\int\limits_{\TT^*}
  \|\fh (z,\cdot)\|_{_{L^2(\W)}}^2\dn z,
\end{equation}
where $\dn k$ is the normalized to total measure $1$ Lebesgue measure on $\B$, and $\dn z$ is the normalized
Haar measure on $\TT^*$.
\item For any $a \in (0,\infty]$, Bloch-Floquet transform
$$
f \mapsto \fh
$$
is a topological isomorphism of the space $L^2_a(\RR^n)$ onto
$A(\Omega_a,L^2(\mathbf{W}))$.
\item
For any $f \in L^2(\RR^n)$ the following inversion
formula holds:
\begin{equation}\label{E:Gelf_inversion}
f(x)=\int\limits_{\TT^*} \fh (k,x) \dn k, \,\, x \in \RR^n.
\end{equation}
\end{enumerate}
\end{theorem}
The first statement claims that the Bloch-Floquet transform is an isometry between the natural Hilbert spaces, the second shows that exponential decay transforms into analyticity in a neighborhood of the torus $\TT^*$ (a Paley-Wiener type theorem), and the third one provides inversion formulas for the transform.

Due to periodicity of the eigenfunction equation $Lu=\lambda u$, the Floquet transform block diagonalizes it. This leads to the well known (see \cite{ReedSimon,AM,Kuch_book,Kuch_UMN} and references there) description of the band-gap spectral structure of the operator $L$. Namely, let $L(k)$ be the operator $L$ acting on the Bloch functions (\ref{E:Bloch}) with a fixed quasi-momentum $k$. If we denote (for real $k$) the eigenvalues of $L$ in nondecreasing order as
\begin{equation}\label{E:fl_spectrum}
    \lambda_1(k)\leq \lambda_2(k)\leq \dots \mapsto \infty,
\end{equation}
then
\begin{enumerate}
  \item \textbf{band functions} $\lambda_j(k)$ are continuous, $\G^*$-periodic, and piece-wise analytic in $k$;
  \item if $I_j$ is the finite segment that is the range of $\lambda_j(k)$ (the \textbf{$j$th band}), then the spectrum of $L$ is
  $$
  \sigma(L)=\bigcup\limits_j I_j.
  $$
\end{enumerate}
Analytic properties of the band functions and the corresponding \textbf{Bloch eigenfunctions} $\phi_j(k,x)$ (i.e., $L\phi_j=\lambda_j\phi_j$) are considered in detail in \cite{Kuch_book,Wilcox}. In particular, one can always choose $\phi_j(k,\cdot)$ as a piece-wise analytic $L^2$-function on $\B$ with values in $L^2(\W)$, whose norm in $L^2(\W)$ is almost everywhere constant and can be chosen equal to $1$. We will also assume below that $\phi(k,\cdot)$ is $\G^*$-periodic with respect to $k$.

For constant coefficient operators, the basis of delta functions is dual under Fourier transform to the basis of plane waves and is also very useful. An attempt of inventing in this vein an appropriate for periodic problems analog of delta functions immediately leads to the so called Wannier functions.

\section{Wannier functions}\label{S:wannier}
Let $\phi_j(k,x)$ be a Bloch eigenfunction (not necessarily normalized) corresponding to the band function $\lambda_j(k)$ (the choice of such an eigenfunction is non-unique, even when the eigenvalue is simple). We will assume that $\phi_j\in L^2(\TT^*,L^2(\W))$, which is known to be always possible (\cite{Kuch_book,Wilcox} and references there).

\begin{definition}
The {\em Wannier function} $w_j(x)$ corresponding to the Bloch eigenfunction  $\phi_j(k,x)$ is
\begin{equation}\label{E:wannier}
    w_j(x)=\int\limits_{\T^*} \phi_j(k,x) \dn k,\quad x\in\RR^n.
\end{equation}
\end{definition}

Comparing this definition with (\ref{E:Gelf_inversion}), one concludes that the Wannier function $w_j$ is just the inverse Bloch-Floquet transform of $\phi_j(k,x)$, and correspondingly $\phi_j(k,x)$ is the Bloch-Floquet transform of $w_j$:
\begin{equation}\label{E:wan_Floquet}
    \phi_j(k,x)=\widehat{w_j}(k,x)=\sum\limits_{\g\in\G}w_j(x+\g)e^{-ik\cdot\g}.
\end{equation}
The analog of this property for the usual Fourier analysis is that plane waves can be obtained as Fourier transforms of delta functions.

Equality (\ref{E:wan_Floquet}) enables one to rephrase the definition of a Wannier function as follows:\\
{\em $w(x)$ is a Wannier function for a periodic operator $L$, if its Bloch-Floquet transform $\widehat{w}(k,x)$ for any $k\in\RR^n$ is an eigenfunction of the corresponding Floquet operator $L(k)$.}

It is often useful to have a Wannier function $w_j(x)$ with mutually orthogonal lattice shifts $w_j(x-\g)$. In fact, these shifts are often considered as different Wannier functions.  It is immediate to check that
\begin{equation}\label{E:wannier_shift}
    w_j(x+\g)=\int\limits_\B e^{ik\cdot\g}\phi_j(k,x) \dn k \mbox{ for }\g\in\G.
\end{equation}
One also wants to have normalized Wannier functions. In these directions, simple answers are provided by the following corollary of (\ref{E:shift}) and Theorem \ref{T:Planch}:
\begin{corollary}\label{C:wannier_orthog}
\indent
\begin{enumerate}
 \item
The Wannier function $w_j(x)$ belongs to $L^2(\RR^n)$ and
\begin{equation}\label{E:wannier_l2}
    \int\limits_{\RR^n}|w_j(x)|^2dx=\int\limits_{\B} \|\phi_j(k,\cdot)\|^2_{L^2(\W)} \dn k.
\end{equation}
\item Functions $w_{j,\g}(x):=w_j(x-\g)$ are mutually orthogonal for $\g\in\G$ if and only if the eigenfunction $\phi_j(k,x)$ in (\ref{E:wannier}) has a $k$-independent norm in $L^2(\W)$.
\end{enumerate}
\end{corollary}
The first claim follows from the first statement of Theorem \ref{T:Planch}. The second one follows from (\ref{E:wannier_shift}) and the fact that the only periodic functions on $\B$ that are orthogonal to all exponents $e^{ik\cdot\g}$ with $\g\in\G,\g\neq 0$, are constants.

If the family $\phi_j(k,x)$ is not normalized, then the shifts of the corresponding Wannier function are not mutually orthogonal. However, if the norm $\|\phi_j(k,\cdot)\|_{L^2(\W)}$ never vanishes, one can normalize it to
$$
\psi_j(k,\cdot)=\frac{\phi_j(k,\cdot)}{\|\phi_j(k,\cdot)\|_{L^2(\W)}}
$$
to get a new Wannier function
$$
w_j^\prime(x)=\int\limits_\B \psi_j(k,x) \dn k.
$$
that already has orthogonal shifts.

The most important property of Wannier functions in many problems of physics is their decay. The already established square integrability is far from being sufficient, and exponential decay is often desired.

Considering the Bloch eigenfunction $\phi(k,x)$ as a vector-valued function of the quasi-momentum $k$, it is clear (as a direct consequence of Theorem \ref{T:Planch}) that smoothness of $\phi_j(k,x)$ with respect to $k$ translates into decay of $w_j(x)$. (Since periodicity of $\phi_j(k)$ is assumed, smoothness also means matching the values and all derivatives across the boundaries of Brillouin zones.) In particular, the following simple statement holds (see \cite[Section 2.2]{Kuch_book} and Theorem \ref{T:Planch} above):
\begin{lemma}\label{L:smooth_wannier}
\indent
\begin{enumerate}
\item If $\sum\limits_{\g\in\G}\|w_j\|_{L^2(\W+\g)}<\infty$, then $\phi_j(k,\cdot )$ is a continuous $L^2(\W)$-valued function on $\T^*$.
\item Infinite differentiability of $\phi_j(k,\cdot )$ as a function on $\T^*$ with values in $L^2(\W)$ is equivalent to the decay of $\|w_j\|_{L^2(\W+\g)}$ faster than any power of $|\g|$ for $\g\in\G$.
\item Analyticity  of $\phi_j(k,\cdot )$ as a function on $\T^*$ with values in $L^2(\W)$ is equivalent to the exponential decay of $\|w_j\|_{L^2(\W+\g)}$.
\end{enumerate}
\end{lemma}
Theorem \ref{T:Planch} provides more details about the relation between the decay and analyticity domain in the last statement of the Lemma.

Thus, it is important to address the issue of choosing $\phi_j(k,\cdot)$ as smooth with respect to $k$ as possible.
We will look at this problem first for a single spectral band.

\section{A single band case}\label{S:singleband}
Dealing with a single band $\lambda_j(k)$ and the eigenfunction branch  $\phi_j(k,\cdot)$, one encounters (at least in the multi-dimensional case, unlike $1D$ \cite{Kohn}) a problem whenever band functions cross, since then smooth continuation of the eigenfunction and the eigenvalue through the crossing might not be (and usually is not) possible \cite{ZK}. Thus, one should assume a band that does not intersect other bands. In this case, although the band function $\lambda_j(k)$ itself is periodic and analytic with respect to $k$ in a neighborhood of the real space, choosing an analytic branch of the eigenfunction that never vanishes might still be impossible. In fact, even continuity of this function might not be achievable. Indeed, finding an analytic and periodic $\phi_j(k,\cdot)$ such that the norm $\|\phi_j(k,\cdot)\|_{L^2(\W)}$ never vanishes, means finding a non-vanishing periodic analytic section of the one-dimensional periodic analytic complex vector bundle over $\RR^n$:
\begin{equation}\label{E:ker_bundle}
   \Lambda_j:= \mathop{\bigcup}\limits_{k\in\RR^n}\ker(L(k)-\lambda_j(k)I)
\end{equation}
Here $\ker A$ means the kernel of the operator $A$, i.e. the space of solutions $Au=0$.

Taking into account $\G^*$-periodicity and considering Floquet multipliers $z=e^{ik}$ instead of quasimomenta, one can reinterpret this bundle as one over the torus $\T^*$. The latter view is the one we will adopt here, while keeping the same notation for the bundle.

Due to one-dimensionality of the bundle, existence of a continuous non-vanishing section over $\T^*$ of $\Lambda_j$ is equivalent to its triviality. The bundle, however, might be non-trivial, which will be a topological obstruction to existence of not only analytic, but even continuous non-vanishing section.

One might wonder whether existence of an analytic section (and thus of an exponentially decaying Wannier function with orthogonal shifts) faces more obstructions than just the existence of a continuous one (which would lead to a much slower decaying Wannier function with orthogonal shifts). As the following proposition states, this happens to be not the case\footnote{This statement has been well known and is implicitly present in \cite{Nenciu,Thouless}.}.
Before formulating it, let us notice first that if $a>0$ is small enough, the band function $\lambda_j(k)$ extends analytically into the neighborhood $\D_a$ of the real space $\RR^n\subset\CC^n$ (or to its periodized version $\Omega_a$, as we will implicitly assume) without colliding with other band functions. Then the bundle $\Lambda_j$ naturally extends analytically to $\Omega_a$:
\begin{equation}\label{E:ker_bundle_compl}
   \Lambda_{j,a}:= \mathop{\bigcup}\limits_{z\in\Omega_a}\ker(L(k)-\lambda_j(k)I), z=e^{ik}.
\end{equation}
\bp \label{P:Oka}
The bundle $\Lambda_{j}$ over $\T^*$ is topologically trivial if and only if for a small $a>0$ the bundle $\Lambda_{j,a}$
over $\Omega_a$ is analytically trivial.
\ep
\proof
We notice first of all that $\Omega_a$ is a product of one-dimensional complex domains, and thus is a Stein manifold (e.g., \cite{Gunn}). Then a Grauert theorem (an incarnation of ``Oka's principle'') applies that says that topological and analytic triviality of $\Lambda_{j,a}$ are equivalent \cite{Grauert1,Grauert2,Grauert3}. Since $\T^*$ is a deformation retract of $\Omega_a$, the triviality of $\Lambda_{j,a}$ over $\Omega_a$ is equivalent to triviality of $\Lambda_{j}$ over $\T^*$.
\eproof

Thus, one needs to worry about topological obstructions only.
There is one important case when these obstructions do not materialize. This is when the operator $L$ has real coefficients. Then the complex conjugate $\overline{u}$ to any solution $u$ of the equation $Lu=\lambda u$ with real $\lambda$ is also a solution. In particular, if $u(x)=e^{ik\cdot x}v(x)$ is a Floquet solution with the quasimomentum $k$, then $\overline{u(x)}=e^{-ik\cdot x}\overline{v(x)}$ is a Floquet solution with the quasimomentum $-k$. Thus, in the Floquet structure of $L$ the symmetry $k\mapsto -k$ is present. In physics this corresponds to the {\em time reversal symmetry} and holds for Schr\"{o}dinger operators with real electric potentials, but breaks down in the presence of a magnetic field (since it contributes imaginary first order terms to the operator).

The following result is proven in \cite{Nenciu}:
\bt\cite{Nenciu}\label{T:nenciu}
If the coefficients of the self-adjoint elliptic $\G$-periodic operator $L$ are real and $\lambda_j(k)$ is an analytic band function without crossings (for real $k$) with other band functions, then the bundle $\Lambda_{j,a}$ (defined for small $a>0$) is analytically trivial. Hence, there exists a corresponding exponentially decaying Wannier function $w_j(x)$ whose $\G$-shifts $w_{j,\g}(x)=w_j(x-\g)$ are mutually orthogonal and normalized in $L^2(\RR^n)$.
\et

If the band function $\lambda_j$ does not have crossings with other band functions, in dimensions $n>1$ the corresponding band $I_j$ of the spectrum still can overlap with other bands. Assume now that this does not happen and the band $I_j=[a_j,b_j]$ is surrounded by gaps separating it from the rest of the spectrum. Then there is a decomposition of the space $L^2(\RR^n)$ into the orthogonal sum of $L$-invariant subspaces
$$
L^2(\RR^n)=H_j\bigoplus H_j^\perp,
$$
such that the spectrum of $L$ in $H_j$ coincides with the band $I_j$, while the spectrum of $L_{H_j^\perp}$ is the rest of the spectrum of $L$. In other words, $H_j$ is the spectral subspace of $L$ corresponding to $I_j$. Then Theorem \ref{T:nenciu} leads to
\bc\label{C:nenciu} \cite{Nenciu}
In the presence of time reversal symmetry, if the spectral band $I_j$ does not intersect other bands of the spectrum of $L$, then there exists an exponentially decaying Wannier function $w_j(x)$ such that its $\G$-shifts $w_{j,\g}(x)=w_j(x-\g)$ form an orthonormal basis in $H_j$.
\ec
\proof
According to Theorem \ref{T:nenciu}, there exists an analytic, $\G^*$-periodic in $k$, and normalized function $\phi_j(k,\cdot)\in Ker(L(k)-\lambda_j(k)I)$. According to Corollary \ref{C:wannier_orthog}, its inverse Bloch-Floquet transform gives a Wannier function $w_j$ with normalized and orthogonal $\G$-shifts $w_{j,\g}$. It only remains to show that these functions are complete in the spectral subspace $H_j$ to which they all belong. Indeed, after Bloch-Floquet transform, the subspace $H_j$ becomes the space of all vector functions of the form $f(k)\phi(k,\cdot)$, where $f(k)$ is any $\G^*$-periodic scalar function such that $\int\limits_\B\|f(k)\|^2\dn k<\infty$. On the other hand, finite linear combinations of Wannier functions $\sum\limits_\G \alpha_\g w_{j,\g}$ transform into functions $\left(\sum\limits_\G \alpha_\g e^{ik\cdot\g}\right)\phi(k,\cdot)$. Since, according to the standard Fourier series theory \cite{Stein}, trigonometric polynomials $\left(\sum\limits_\G \alpha_\g e^{ik\cdot\g}\right)$ are $\G^*$-periodic and dense in $L^2(\B)$, this proves completeness of the set of Wannier functions $w_{j,\g}$ in $H_j$.
\eproof

\section{Multiple bands and composite Wannier functions}\label{S:multiband}
Suppose now that a set $S$ consists of $m>1$ bands of the spectrum (with overlaps and crossings allowed) and is separated from the rest of the spectrum by spectral gaps. An example would be the part of the spectrum from its bottom till the first gap. Let us denote the union of these bands by $S$ (a {\bf composite band}).
A natural question to ask now is the following: are there analogs of Theorem \ref{T:nenciu} and Corollary \ref{C:nenciu} for this case? Again, there exists an orthogonal decomposition into $L$-invariant subspaces $H_S\bigoplus H_S^\perp$, where $H_S$ is the spectral subspace of $L$ corresponding to the union of bands $S$.
After the Floquet transform, functions $f\in H_S$ will correspond to functions $\widehat{f}(k,\cdot)$ that for each $k\in\B$ belong to the spectral subspace $H_{S,k}\in L^2(\W)$ of the Floquet operator $L(k)$ that corresponds to the set $S$.

Since the band functions that correspond to the bands in $S$ might cross, the attempts to find analytic non-degenerate families of eigenfunctions $\phi (k,x)$ are expected to be futile in general. However, one can relax the requirement
that $\phi (k,x)$ is an eigenfunction of $L(k)$ and request only that $\phi (k,\cdot)\in H_{S,k}$ for each real $k$.
This leads to the following notion:

\bd\label{D:wann_general}
A {\bf composite (or generalized) Wannier function} $w(x)$ corresponding to an isolated from the rest of the spectrum union $S$ of bands is the one that can be represented as
\begin{equation}\label{E:wannier}
    w(x)=\int\limits_{\T^*} \phi(k,x) \dn k,
\end{equation}
where $\phi (k,\cdot)\in H_{S,k}$ for all real $k$.
\ed
Let us assume that $S$ consists of $m$ bands. One can ask the question: {\em do there exist $m$ exponentially decaying generalized Wannier functions $w_j$ such that their shifts $w_{j,\g}(x)=w_j(x-\g),\g\in\G$ form an orthonormal basis in the spectral subspace $H_S$ of the operator $L$?}

In order to attempt an answer, let us consider a simple contour $\mathcal{C}$ in the complex plane that surrounds the set $S$ and separates it from the rest of the spectrum of $L$. It does not intersect the spectrum of $L(k)$ for any real $k$, and thus, by a perturbation argument, does not intersect it for all quasimomenta $k$ with sufficiently small imaginary parts. Hence, if one considers the $m$-dimensional spectral projector $P_S(k)$ for $L(k)$ in $L^2(\W)$ that corresponds to this contour
\begin{equation}\label{E:projector}
P_S(k)=\frac{1}{2\pi i}\oint\limits_{\mathcal{C}}(\zeta I-L(k))^{-1}d\zeta,
\end{equation}
the projector will be analytic with respect to $k$ in a ($\G^*$-periodic) neighborhood of the space of real quasimomenta $k$. Thus, its range forms an analytic $\G^*$-periodic vector bundle
\begin{equation}\label{E:bundle}
\Lambda_S=\mathop{\bigcup}\limits_{\D_a} P_S(k)(L^2(\W))
\end{equation}
over a neighborhood $\D_a$ of the real space. As before, taking into account periodicity and considering Floquet multipliers $z=e^{ik}$ instead of quasimomenta, one can reinterpret this bundle as one over a neighborhood $\Omega_a$ of the unit torus $\T^*$. The latter view is the one we will adopt here, without changing notations for the bundle.

Taking into account our previous discussion and the requirement that the number of the ``mother'' Wannier functions $w_j$ is assumed to be equal to the number $m$ of bands in $S$, one obtains the following

\bl \cite{Panati}\label{L:gen_wannier}
Let $S$ be the union of $m$ bands that is isolated from the rest of the spectrum. A family of $m$ exponentially decaying generalized Wannier functions $w_j$ such that their shifts $w_{j,\g}(x)=w_j(x-\g),\g\in\G$ form an orthonormal basis in the spectral subspace $H_S$ of the operator $L$ exists
if and only if the bundle $\Lambda_S$ over $\Omega_a$ is analytically trivial. Equivalently (due to Oka's principle), this happens if and only if the restriction of $\Lambda_S$ to $\T^*$ is topologically trivial.
\el

The following complement to Theorem \ref{T:nenciu} requires non-trivial topological arguments and was proven in \cite{Panati}:

\bt\cite{Panati}\label{T:panati}
Let the coefficients of the self-adjoint elliptic $\G$-periodic operator $L$ in $\RR^n, n\leq 3$ be real (and thus the time reversal symmetry holds) and $S\subset\RR$ be the union of $m$ spectral bands of $L$. We assume that $S$ is separated by gaps from the rest of the spectrum. Then the bundle $\Lambda_S$ is analytically trivial. Hence, there exist $m$ exponentially decaying generalized Wannier functions $w_j(x)$ whose $\G$-shifts $w_{j,\g}(x)=w_j(x-\g)$ form an orthonormal basis in $H_S$.
\et

The nice Theorems \ref{T:nenciu} and \ref{T:panati} give the desired positive answer in physical dimensions $n\leq 3$ and any $m$ and for any dimension $n$ when $m=1$, whenever the time reversal symmetry is present. However, without this symmetry (or in dimensions $4$ and higher, when $m>1$) one cannot guarantee triviality of the bundle and thus existence of the required generalized Wannier basis in the spectral subspace $S$. One can wonder what can be achieved without triviality of the bundle. According to Lemma \ref{L:gen_wannier}, if the bundle $\Lambda_S$ is non-trivial, one should sacrifice some of the properties guaranteed in Theorem \ref{T:panati}: exponential decay of generalized Wannier functions, their number (i.e., one might need more Wannier functions than the number $m$ of the bands),  orthogonal basis property, or several of these.
One might try to weaken the exponential decay condition to avoid redundancy in the system of Wannier functions and thus to keep the basis property. The next simple statement shows that in the presence of the topological obstacle, one cannot go far in this direction, since even the weak decay condition (\ref{E:slow}) below is already as strong as the exponential one.
\begin{theorem}\label{T:slow}
Let $L$ be a self-adjoint elliptic $\G$-periodic operator in $\RR^n, n\geq 1$, and $S\subset\RR$ be the union of $m$ spectral bands of $L$. We assume that $S$ is separated from the rest of the spectrum. Suppose that there exists a family of $m$ composite Wannier functions $w_j(x)$ whose $\G$-shifts form an orthonormal basis in $H_S$ and such that for each $j=1,\dots,m$ the following sum is finite
\begin{equation}\label{E:slow}
\sum\limits_{\g\in\G}\|w_j\|_{L^2(\W+\g)} < \infty.
\end{equation}
Then the bundle $\Lambda_S$ is trivial and thus there exists a system of $m$ exponentially decaying Wannier functions $v_j$ that have the same basis property. In particular, power decay as
$$
\|w_j\|_{L^2(\W+\g)}\leq C (1+|\g|)^{-n-\epsilon}
$$
for some $\epsilon>0$ (or a stronger requirement that $|w(x)|\leq C (1+|x|)^{-n-\epsilon}$)
already implies absence of the topological obstacle.
\end{theorem}
\proof
Indeed, according to Lemma \ref{L:smooth_wannier}, condition (\ref{E:slow}) implies continuity with respect to $k$ of the corresponding Bloch functions $\phi_j(k,\cdot)$. Together with orthonormality condition, this gives a basis of continuous sections, and thus triviality of $\Lambda_S$. This, in turn, implies, as we have already seen, existence of the corresponding family of exponentially decaying Wannier functions. \eproof

So, we would like to preserve the exponential decay. The next statement shows that one can do so at the cost of increasing the number of Wannier functions and thus sacrificing their orthogonality, while still having a complete system of functions. This is a preliminary simple result, since Theorem \ref{T:wan_kuch_frame} proved later improves significantly on the completeness statement and provides an estimate of redundancy of the system of Wannier functions.

\bt\label{T:wan_kuch}
Let $L$ be a self-adjoint elliptic $\G$-periodic operator in $\RR^n, n\geq 1$, and $S\subset\RR$ be the union of $m$ spectral bands of $L$. We assume that $S$ is separated from the rest of the spectrum. Then there exists a finite number $l\geq m$ of exponentially decaying composite Wannier functions
$w_j(x)$ such that their shifts $w_{j,\g},\g\in\G$ form a complete system in the spectral subspace $H_S$ (i.e., the set of all their finite linear combinations is dense in the space).
\et
\proof We will prove this statement in an equivalent formulation: there exists a finite number of analytic sections $\phi_j(k,\cdot),j=1,\dots,l$ of the bundle $\Lambda_S$ over $\Omega_a$ for some $a>0$ such that the sections $\phi_{j,k}(k,\cdot):=e^{ik\cdot \g}\phi_j(k,\cdot)$ for $j=1,\dots,l,\g\in\G$ form a complete set of functions in the space of $L^2$-sections of the bundle $\Lambda_S$. It is a simple exercise to show that it is sufficient for that to have a finite family of analytic sections $\phi_j$ such that they span the whole fiber of the bundle at each point $z\in\T^*$. Recall now that the bundle $\Lambda_S$ is trivial locally, i.e. in a neighborhood of each point. Let us fix two finite trivializing covers $U_j$ and $V_j\Subset U_j, j=1,\dots,r$ of $\T^*$ by complex domains such that their union is a  Stein domain (e.g., the union can be made equal to $\Omega_a$ for some $a>0$). Then on each $U_j$, by assumption, there is an analytic basis $\psi_{j,t}(k), t=1,\dots,m$ in the restriction of the bundle $\Lambda_S$ onto $U_j$. Due to the Stein property, these holomorphic sections can be approximated with any precision uniformly on $\overline{V_j}$ by global (i.e., defined on the whole $\Omega_a$) analytic sections $\phi_{j,t}(k)$ (see, e.g. \cite[Ch. VII]{Horm_compl} or \cite[Theorem 1.5.9]{Kuch_book}). Doing this for each $j$, with a sufficiently close approximation, one achieves the desired property for the functions $\phi_{j,t}$.
\eproof

In this result we seemingly have lost too much: we have an excessive (in comparison with the number of spectral bands involved) and uncontrollable number of composite Wannier functions, and besides the orthonormal property is lost, being replaced by a much weaker (and not very useful) completeness. Well, one cannot avoid having more than $m$ generalized Wannier functions. Indeed, the following statement holds:

\bp\label{P:wan_excess}
If in the previous theorem either the number of Wannier functions is exactly equal to $m$, or the shifts of the Wannier functions $w_j$ form an orthonormal basis, then the bundle $\Lambda_S$ is trivial.
\ep
\proof \indent
Indeed, assume that the number of the functions is exactly $m$. It is not hard to notice that in this case the Floquet transforms $\phi_j$ of the Wannier functions must be analytic and at each point must span the whole fiber of the bundle. Since the fibers have dimension $m$, we conclude that $\phi_j$ must form an analytic basis of the bundle, and thus the bundle is trivial.

Suppose now that the functions $w_{j,\g}$ are orthonormal. Normalization property, as we already know, means that the functions $\phi_j(k,\cdot)$ have a constant norm in $L^2(\W)$. Checking orthogonality of $w_{j_1,0}$ to $w_{j_2,\g}$ for $j_1\neq j_2$ and arbitrary $\g\in\G$, one obtains that
$$
\int\limits_{\T^*}e^{ik\cdot\g}(\phi_{j_1}(k,\cdot),\phi_{j_2}(k,\cdot))_{L^2(\W)}dk=0
$$
for all $\g\in\G$. This means that the whole Fourier series with respect to $k$ of the function $(\phi_{j_1}(k,\cdot),\phi_{j_2}(k,\cdot))_{L^2(\W)}$ on $\T^*$ vanishes. Thus, $(\phi_{j_1}(k,\cdot),\phi_{j_2}(k,\cdot))_{L^2(\W)}$ is equal to zero almost everywhere, and hence the functions  $\phi_{j_1},\phi_{j_2}$ are orthogonal for each $k$. Since the dimension of the fiber is $m$, we conclude that the number of the Wannier functions does not exceed $m$, and thus, according to the previous case, the bundle is trivial.
\eproof

So, if one cannot help with excessive number of Wannier functions, can one improve on the completeness statement?
It seems that the answer must be negative, since an overdetermined system of vectors cannot be orthogonal. However, we have described in Definition \ref{D:frame} the notion of the so called {\bf tight frame} of vectors (e.g., \cite{Larson_frame}) that replaces orthonormality in the redundant case. This is what allows us now to improve on the statement of Theorem \ref{T:wan_kuch} and prove our main result.

\bt\label{T:wan_kuch_frame}
Let $L$ be a self-adjoint elliptic $\G$-periodic operator in $\RR^n, n\geq 1$ and $S\subset\RR$ be the union of $m$ spectral bands of $L$. We assume that $S$ is separated from the rest of the spectrum. Let also $\tau$ be the type of the obstacle bundle $\Lambda_S$ over the $n$-dimensional torus $\TT^*$. Then there exists a finite number $l\leq\tau m$ (and thus $m\leq l \leq 2^n m$) of exponentially decaying composite Wannier functions
$w_j(x)$ such that their shifts $w_{j,\g}:=w_j(x-\g),\g\in\G$ form a $1$-tight frame in the spectral subspace $H_S$ of the operator $L$. This means that for any $f(x)\in H_S$, the equality holds
\begin{equation}
    \int\limits_{\RR^n}|f(x)|^2dx=\sum\limits_{j,\g}|\int\limits_{\RR^n}f(x)\overline{w_{j,\g}(x)}dx|^2.
\end{equation}
The number $l\in [m,2^n m]$ is the smallest dimension of a trivial bundle containing an equivalent copy of $\Lambda_S$. In particular, $l=m$ if and only if $\Lambda_S$ is trivial, in which case an orthonormal basis of exponentially decaying composite Wannier functions exists.
\et
\proof
Let us establish first of all the following auxiliary statement:

\bl\label{L:bundle_complement}
There exists a finite dimensional analytic sub-bundle $\Lambda^\prime$ of the trivial bundle $\Omega_a\times L^2(\W)$, such that
\begin{enumerate}
  \item The fibers at the same point $k$ of $\Lambda_S$ and $\Lambda^\prime$ intersect only at zero.
  \item The analytic sub-bundle $\Phi:=\Lambda_S\bigoplus\Lambda^\prime$ in  $\Omega_a\times L^2(\W)$ is trivial and has dimension $l$, with $m\leq l \leq 2^n m$ as described in Theorem \ref{T:wan_kuch_frame}.
\end{enumerate}
\el
There are several ways to show that this lemma holds. For instance, one way is to choose an abstract finite-dimensional analytic bundle $\Lambda^\prime$ such that added to  $\Lambda_S$ restricted to some $\overline{\Omega_b}$ (where $\overline{\Omega}$ is the closure of $\Omega$) it leads to a trivial bundle of dimension $l$.
This is possible in the topological category \cite{Atiyah,Husemoller} and thus, according to Grauert's theorem \cite{Grauert3} and Stein property of $\Omega_a$, also in analytic category. Using Gaussian maps, one can make sure (e.g., proposition 3.5.8 in \cite{Husemoller}) that the dimension $l$ of the direct sum $\Lambda_S\bigoplus \Lambda^\prime$ does not exceed $\tau m$, where $m$ is the dimension of the fibers of the bundle $\Lambda_S$ and $\tau$ is its type (see Definition \ref{D:type}). In the case of an $n$-dimensional torus as the base, as it is mentioned in Remark \ref{R:type}, $\tau$ does not exceed $2^n$. Thus, one can always achieve $l\leq 2^n m$.

Now let us consider a complementary to $\Lambda_S$ infinite dimensional sub-bundle $\mathcal{F}$ in $\Omega_a\times L^2(\W)$, which always exists \cite[Theorem 3.11]{ZK} (and in our case can be constructed using the spectral projectors, as in (\ref{E:projector})). The claim is now that there is an analytic sub-bundle in $\mathcal{F}$ that is isomorphic to $\Lambda^\prime$. Indeed, taking the direct sum of $\Lambda^\prime$ and of the trivial bundle with an infinite dimensional separable Hilbert space $G$ as a fiber, one gets a Hilbert analytic fiber-bundle over a Stein manifold. According to Kuiper's theorem \cite{Atiyah,Kuiper}, this bundle is topologically trivial. Bungart's theorem \cite{Bungart} (an infinite dimensional version of Grauert's theorem \cite{Grauert3} on fiber bundles on Stein spaces) implies that this sum is also analytically trivial. Since, again by the Kuiper and Bungart theorems, all analytic infinite dimensional Hilbert bundles with separable fibers over Stein bases are equivalent, we conclude that the direct sum is equivalent to $\mathcal{F}$, which embeds a copy of $\Lambda^\prime$ into $\mathcal{F}$.  This proves the Lemma.

Let us now prove the statement of the theorem, using the sub-bundle $\Lambda^\prime$ as a tool. Consider the trivial $l$-dimensional bundle $\Phi:=\Lambda_S\bigoplus\Lambda^\prime\subset \Omega_a\times L^2(\W)$. Due to its triviality, one can find a basis of its analytic sections $\psi_j(k),j=1,\dots,l$. What we need to do is to make this system orthonormal for real values of $k$ (or equivalently for $z\in\T^*$) without losing the analyticity. This is possible in our case. Indeed, one would like to use the standard Gram-Schmidt orthogonalization method, but in a version that preserves analyticity with respect to the quasi-momentum $k$ (equivalently, with respect to the multiplier $z$). Consider the domain $\D_a$, which is a neighborhood in $\CC^n$ of the space of real quasimomenta $\RR^n$. It is clearly symmetric with respect to the complex conjugation $k=(k_1,\dots,k_n)\mapsto \overline{k}=(\overline{k_1},\dots,\overline{k_n})$. Let $\phi(k)$ be an analytic function in $\D_a$ with values in a complex Hilbert space $H$ ($H=L^2(W)$ will be our case) with the Hermitian scalar product $(\cdot,\cdot)$ (which is thus anti-linear with respect to the second factor). Then the function $(\phi(k),\phi(\overline{k}))$ is analytic in $\D_a$ and coincides with $\|\phi(k)\|^2$ for real $k$. If now $\|\phi(k)\|$ does not vanish for real $k$, then one can normalize $\phi(k)$ as follows:
\begin{equation}\label{E:normal}
    \psi(k):=\frac{\phi(k)}{\sqrt{(\phi(k),\phi(\overline{k}))}}.
\end{equation}
Then $\psi(k)$ is analytic in a neighborhood of the real space and has unit norm for real values of $k$. Notice that the expression in (\ref{E:normal}) can lose its analyticity at some point, due to the denominator, and thus the neighborhood $\D_a$ might need to shrink somewhat. An analogous analytic procedure can be applied to inner products by computing $(\phi(k),\psi(\overline{k}))$. Now this allows one to apply the Gram-Schmidt orthonormalization for real $k$ in a manner that preserves analyticity in a complex neighborhood of $\RR^n$. The same procedure applies in a neighborhood $\Omega_a$ of the torus $\T^*$. We can thus produce another analytic basis family in $\Phi$, which is orthonormal for real values of $k$. We can assume now that $\psi_j(k)$ is already like that. Applying the spectral projector $P_S(k)$ to this family of sections, we obtain an analytic family of sections $\phi_j(k)$ of the bundle $\Lambda_S$, which in each fiber form a $1$-tight frame (as the orthogonal projection of an orthonormal basis in a larger space). Consider the two families of generalized Wannier functions: $v_j$ that correspond to $\psi_j$ and $w_j$ that correspond to $\phi_j$. The functions $v_{j,\g}$ do not necessarily belong to the spectral subspace $H_S$, but according to Theorem \ref{T:Planch} and Corollary \ref{C:wannier_orthog}, they form an orthonormal basis in a Hilbert subspace $H^\prime$ that is larger than $H_S$. (This is the subspace that after the Bloch-Floquet transform produces the space of all $L^2$-sections of the bundle $\Phi$ over the torus.) Thus, the family $w_{j,\g}$, as the orthogonal projection of $v_{j,\g}$ from $H^\prime$ onto $H_S$, form a tight frame in $H_S$.
\eproof

\section{Conclusion}
\begin{enumerate}
\item We have shown that for any isolated finite part $S$ of the spectrum of an elliptic self-adjoint operator periodic with respect to a lattice $\G$, there exists a family of $l$ exponentially decaying composite (generalized) Wannier functions such that their $\G$-shifts form a tight frame in the spectral subspace $H_S$ that corresponds to $S$. The number $l$ is equal to the smallest possible dimension of the fiber of a trivial vector bundle containing the obstacle bundle $\Lambda_S$. In particular, it does not exceed $\tau m$, where $\tau$ is the type of the bundle $\Lambda_S$ and $m$ is the number of spectral bands constituting $S$. Thus, $m\leq l \leq 2^n m$. The numbers $l$ and $m$ are equal, and thus there is no redundancy in the constructed set of Wannier functions, if and only if  $\Lambda_S$ is trivial, and then the shifts of these Wannier functions form an orthonormal basis in $H_S$.

\item It was also shown that an attempt to relax the decay condition in presence of the topological obstruction does not work well, since even a very slow decay condition (\ref{E:slow}) hits  the same topological obstruction.

  \item The results apply to a general class of elliptic self-adjoint periodic operators, including systems such as Dirac operators and Maxwell operators in periodic media. Magnetic translations can also be treated the same way.

  \item The results also apply in an abstract situation of a periodic elliptic self-adjoint operator on an abelian covering of a compact manifold, graph, or quantum graph. Neither formulations, nor proofs require any modifications in this case.
\end{enumerate}

\section*{Acknowledgments}
The author expresses his gratitude to the NSF grant DMS-0406022  for partial support of this work and to D.~Larson, from whose lectures he has learned about the frame theory. The author is also thankful to the referees for many useful remarks and suggestions.

\end{document}